\documentclass[11pt]{article}
\usepackage{geometry}                
\usepackage[parfill]{parskip}    
\usepackage{graphicx}
\usepackage{amsmath, amsthm, amsfonts}
\usepackage{amssymb}

\usepackage{color}
\usepackage[colorlinks]{hyperref}
\hypersetup{linkcolor=blue,%
citecolor=red,%
urlcolor=cyan}
\usepackage{url} 

\usepackage{epstopdf}
\title{\bf  Dirac-like Hamiltonians associated to Schr\"odinger factorizations}

\author { D. Demir K{\i}z{\i}l{\i}rmak$^a$\footnote{duygudemirkizilirmak@gmail.com, ORCID: \href{https://orcid.org/0000-0002-7381-8327}{0000-0002-7381-8327}},
 \c{S}. Kuru$^b$\footnote{sengul.kuru@science.ankara.edu.tr, ORCID: \href{http://orcid.org/0000-0001-6380-280X}{0000-0001-6380-280X}}, 
J. Negro$^c$\footnote{jnegro@fta.uva.es, ORCID: \href{http://orcid.org/0000-0002-0847-6420}{0000-0002-0847-6420}}   
\bigskip
\\ 
\noindent
$^a$\,Department of Medical Services and Techniques, Ankara Medipol University, 06050 Ankara, Turkey
\\
\noindent
$^b$\,Department of Physics, Faculty of Science, Ankara
University, 06100 Ankara, Turkey
\\ 
 \noindent
$^c$\,Departamento de F\'{\i}sica Te\'orica, At\'omica y
\'Optica, and IMUVA,\\ Universidad de Valladolid,  47011 Valladolid, Spain}

\begin{document}

\maketitle

\begin{abstract}
In this work, we have extended the factorization method of scalar shape-invariant Schr\"o\-din\-ger Hamiltonians to a class of Dirac-like matrix Hamiltonians. The intertwining operators of the Schr\"odinger equations have been implemented in the Dirac-like shape invariant equations. We have considered also another kind of anti-intertwining operators changing the sign of energy. The Dirac-like Hamiltonians can be obtained from reduction of higher dimensional spin systems. Two examples have been worked out, one obtained from the sphere ${\cal S}^2$ and a second one, having a non-Hermitian character, from the hyperbolic space
${\cal H}^2$.
\end{abstract}



\section{Introduction}
\label{intro}

In this work, our starting point will be the one-dimensional Schr\"odinger equations that can be algebraically solved by the well known method of factorization \cite{infeld}. This method has a long history of applications to quantum mechanics since the early times when it was already used by Dirac and Schr\"odinger \cite{dirac,schr}. The basis of the method can be described as the building of a sequence (or hierarchy) of shape invariant Schr\"odinger Hamiltonians connected by intertwining operators \cite{COO,david10,cooper}. The knowledge of such intertwining operators and their annihilated functions solves the eigenvalue problem for all the eigenvalue equations in the Hamiltonian hierarchy.
Our objective here is to widen the factorization method from the Schr\"odinger second order scalar equations to a kind of first order matrix Dirac-like equations.
We will associate to each of these hierarchies of  Schr\"odinger equations another 
hierarchy  made up of matrix equations similar to
Dirac Hamiltonians in one variable \cite{greiner}. This type of Dirac-like factorization hierarchies will have also a kind
of intertwining operators which correspond to those of the scalar case in a natural way. Let us remark that some particular cases of such Dirac-like Hamiltonian hierarchies have been found previously when solving the Dirac equation for some external magnetic fields
\cite{sandro,alonso,david,negro09,Jakubsky13,roy20,schulze,demir20,demir21}. In this work we plan
to undertake a systematic study of such a kind of matrix Hamiltonians.

Among the purposes of this work, we want to 
understand the origin of the ``Dirac factorizations'' from symmetry considerations.
In fact, we will show that the class of Dirac-like Hamiltonians related to  factorizable Schr\"odinger equations can be explained by means of a reduction of higher dimensional problems in flat and curved spaces. These initial systems must have symmetries, including spin, which allow the reducing process.
Our program consist in the following steps. We will begin with a free Hamiltonian on a flat or curved space. If this Hamiltonian is scalar it will lead, by reduction, to some of the Schr\"odinger factorized Hamiltonians. If that Hamiltonian is a kind of spin-orbit coupling then, after a reduction process, it will lead us to a particular kind of 
the Dirac-like Hamiltonians
proposed for each factorization type.
The intertwining operators of the reduced Dirac Hamiltonians will be recovered from the symmetries of the initial system, but we remark that there will appear new intertwining operators. We have restricted in this paper to some specific two-dimensional spaces, the sphere ${\cal S}^2$ and the hyperboloid ${\cal H}^2$, for the sake of simplicity and to keep it within a reasonable length, but this program can in principle be applied to all the factorization cases
\cite{infeld}.

In a similar way as the Klein-Gordon equation is
obtained by ``squaring'' the Dirac equation, we will have a similar property where the scalar second order Schr\"odinger factorized Hamiltonian is obtained by taking the square (in an appropriate way) of the associated Dirac-like Hamiltonian.

The organization of this work is as follows. In the second section, the factorization method is introduced briefly in order to set the notation of increasing and decreasing hierarchies used along the paper. The third section is devoted to  Dirac-like Hamiltonians related with increasing hierarchies. Next, in the fourth section, an example on ${\cal S}^2$ is given and the connection with symmetry group reduction is examined. In the fifth section the same procedure is applied to find 
 Dirac-like Hamiltonians for decreasing hierarchies which in this case are non-Hermitian
 \cite{mosta02}. Here, the space ${\cal H}^2$ will be used
in order to get an example of this type of Dirac-like Hamiltonians. Finally, this work will end with some conclusions and remarks.

\section{Hierarchies of 1D Schr\"odinger Hamiltonians}

In the following sections we will specify how is  the correspondence
of 1D Schr\"odinger and Dirac Hamiltonian hierarchies of equations.
But first, it is necessary to distinguish three cases of hierarchies
of shape invariant Schr\"odinger equations (also referred as
shape-invariant potentials). These are well known \cite{infeld,COO,david10,cooper}, but here we
will briefly introduce them  in order to set the notation.

\begin{itemize}


\item {\bf\em Increasing hierarchy (or I-hierarchy)\/}

Consider the Schr\"odinger hierarchy of second order Hamiltonians $H_n$
which can be factorized as follows:
\begin{equation}\label{ih}
H_n =-\partial_{xx} + V_n(x) = a^+_n a^-_n + \mu_n^2= a^-_{n-1} a^+_{n-1} + \mu_{n-1}^2,\qquad
n=0,1,\dots
\end{equation}
where $\partial_{xx}:=\frac{d^2}{dx^2}$. Let us explain the standard notation here used. The first order operators
$a^\pm_n$ are defined by
\begin{equation}\label{apm}
a^\pm_n(x) = \mp \partial_x + w_n(x)
\end{equation}
Here, $\partial_{x}$ is the derivative with respect to $x$, $\partial_{x}:=\frac{d}{dx}$; the functions $w_n(x)$ are called superpotentials and, according to eq.~(\ref{ih}), satisfy the following relations with the potential $V_n(x)$,
\begin{equation}\label{sp2}
V_n(x) = w_n(x)^2-w_n'(x) +\mu_n^2 =w_{n-1}(x)^2+w_{n-1}'(x)+\mu_{n-1}^2
\end{equation}
where $w_n'(x):=\frac{d w_n(x)}{dx}$. The factorization energies $\mu_n^2$ are assumed to be positive, but the following results can be also adapted to other factorization energies, as we will see later.

In all the hierarchies, in particular in this I-hierarchy, we will
also assume that $a^-_n$ are annihilation operators, in other words
the equation
\begin{equation}\label{psi0}
a^-_n \psi_{n}^0=0,\quad \implies \quad \psi_{n}^0(x) = K\, e^{-\int
w(x) dx}
\end{equation}
(where $K$ is an integration constant) determines a
square-integrable function $\psi_{n}^0(x)$. The function
$\psi_{n}^0(x)$ corresponds to the ground state of $H_n$, while
$\mu_n^2$ is the ground state energy (this is consistent with the
factorization (\ref{ih})):
\begin{equation}\label{e0}
E_{n}^0 = \mu_n^2\,
\end{equation}
This sequence of Hamiltonians $H_n$ has increasing ground energies.
The key factorization property \eqref{ih} implies that the operators
$a^-_n$ intertwine the consecutive Hamiltonians $H_n$ and $H_{n+1}$:
\begin{equation}\label{inti}
a^-_n H_n = H_{n+1} a^-_n,\qquad
 H_n a^+_n = a^+_n H_{n+1}
\end{equation}
The annihilation operators $a^-_n$ go towards the Hamiltonians on
the right (index $n$) but decrease the energy level (index $k$). The
creation operators $a^+_n$ act in the opposite way:
\begin{equation}\label{diramp}
a^-_n: \psi_{n}^k\ \to \ \psi_{n+1}^{k-1},\qquad a^+_n:
\psi_{n+1}^{k-1}\ \to \ \psi_{n}^{k}
\end{equation}
In fact
\begin{equation}
a^-_n\psi_{n}^k=\sqrt{\mu_{n+k}^2-\mu_n^2}\,\psi_{n+1}^{k-1},\qquad
a^+_n\psi_{n+1}^{k-1}=\sqrt{\mu_{n+k}^2-\mu_n^2}\, \psi_{n}^{k}
\end{equation}
The $k$-excited eigenfunction $\psi_{n}^k$ of $H_n$ can be obtained
from the ground eigenfunction $\psi_{n+k}^0$ of $H_{n+k}$, by means
of creation operators:
\begin{equation}\label{enk}
\psi_{n}^k \ \propto \ a^+_n a^+_{n+1}\dots a^+_{n+k-1}
\psi_{n+k}^0\,, \qquad E_{n}^k = \mu_{n+k}^2
\end{equation}
An example of this increasing hierarchy type is the trigonometric P\"oschl-Teller (PT) Hamiltonians as we will see later.

\item {\bf\em Decreasing hierarchy (or D-hierarchy)\/}

Next, we start from a Schr\"odinger hierarchy of second order Hamiltonians $H_n$
satisfying the following factorization property:
\begin{equation}\label{dh}
H_n =-\partial_{xx} + V_n(x) = a^+_n a^-_n - \mu_n^2= a^-_{n+1} a^+_{n+1} - \mu_{n+1}^2,\qquad
n=0,1,\dots
\end{equation}
The first order operators
$a^\pm_n$ are defined in the same way as \eqref{apm}.
The superpotentials $w_n(x)$ and the potentials $V_n(x)$ are here related by
\begin{equation}\label{sp}
V_n(x) = w_n(x)^2-w_n'(x) -\mu_n^2 =w_{n+1}(x)^2+w_{n+1}'(x)-\mu_{n+1}^2
\end{equation}
The factorization energies are given by $-\mu_n^2$, where $\mu_n^2$ are real numbers which we will assume to be positive.
The annihilation operator $a^-_n$
determines  the ground state wavefunction 
$\psi_{n}^0(x)$
of $H_n$,
\[
a^-_n \psi_{n}^0=0,\qquad E_n^0 = -\mu_n^2
\]
where $-\mu_n^2$ is the negative ground state energy, in a similar way as in the previous positive case
(\ref{psi0})-(\ref{e0}). 

Here the sequence of Hamiltonians $H_n$ have
decreasing ground energies.
In the D-hierarchies, the factorization property \eqref{dh} implies that the operators  $a^\pm_n$ intertwine the consecutive Hamiltonians, $H_n$ and $H_{n-1}$, as follows:
\begin{equation}\label{inti2}
a^-_n H_n = H_{n-1} a^-_n,\qquad
 H_n a^+_n = a^+_n H_{n-1}
\end{equation}
The annihilation operators $a^-_n$ go towards the Hamiltonians to
the left (in  index $n$) and decrease the energy level (index $k$).
The creation operators $a^+_n$ go in the opposite way:
\begin{equation}\label{diramp2}
\begin{array}{ll}
a^-_n: \psi_{n}^k\ \to \ \psi_{n-1}^{k-1},\quad
&
a^+_n: \psi_{n-1}^{k-1}\ \to \ \psi_{n}^{k}
\\[2.ex]
a^-_n\psi_{n}^k=\sqrt{\mu_n^2-\mu_{n-k}^2}\,\psi_{n-1}^{k-1},\quad
&
a^+_n\psi_{n-1}^{k-1}= \sqrt{\mu_n^2-\mu_{n-k}^2}\,\psi_{n}^{k}
\end{array}
\end{equation}
The $k$-excited eigenfunction $\psi_{n}^k$ of $H_n$ can be obtained
from the ground eigenfunction $\psi_{n-k}^0$ of Hamiltonians to the left $H_{n-k}$, by means
of creation operators:
\begin{equation}
\psi_{n}^k \ \propto \ a^+_n a^+_{n-1}\dots a^+_{n-k+1} \psi_{n-k}^0,\qquad
E_{n}^k=E_{n-k}^0= -\mu_{n-k}^2,\qquad k=1,2\dots
\end{equation}
An example of this decreasing hierarchy type is the sequence of hyperbolic P\"oschl-Teller Hamiltonians.


\item {\bf\em Equal hierarchy (or E-hierarchy)\/}.

This is a hierarchy where all the factorizacion energies are equal:
\begin{equation}
\mu^2 = \mu_n^2,\qquad \forall n
\end{equation}
In this case the spectrum of all the Hamiltonians $H_n$ are the same (they are
isospectral). The operators $a^\pm_n$ are neither annihilation nor creation, they could be called iso-operators. This case will not be considered in this work.

\end{itemize}
\section{Hierarchies of 1D Dirac 
Hamiltonians: Increasing Case}

Consider an increasing hierarchy of Schr\"odinger Hamiltonians  $H_n$
characterized by the factorization \eqref{ih}. Then, we can form a $2\times 2$ matrix Hamiltonian
hierarchy $h_n$ defined by
\begin{equation}\label{hn1}
h_n = \left(\begin{array}{cc}
\mu_n & i a^+_n\\[1.ex]
-i a^-_n & -\mu_n
\end{array}\right)
\end{equation}
We will refer to this first order matrix operator $h_n$ as a Dirac-like Hamiltonian.
This type of Dirac-like Hamiltonians have been obtained in a number of problems related with Dirac-Weyl systems in planar graphene, spherical surfaces or for other cases with cylindrical symmetry \cite{sandro,alonso,david,negro09,Jakubsky13,roy20,schulze}. 

The matrix Hamiltonian $h_n$ in (\ref{hn1}) is explicitly Hermitian from a formal point of view.
The square of $h_n$ gives a pair of Schr\"odinger Hamiltonians of the scalar I-hierarchy in the diagonal, therefore in this sense it mimics the Dirac Hamiltonians:
\begin{equation}\label{sqr}
h_n^2 = \left(\begin{array}{cc}
H_n & 0\\[1.ex]
0 & H_{n+1}
\end{array}\right) := {\boldsymbol H}_n
\end{equation}
Next, we will look for a pair of first order matrix differential (anti)intertwining operators $A^\pm_n$
between the matrix Hamiltonians $h_n$ and $h_{n+1}$ (similar to the
operators $a^\pm_n$ between $H_n$ and $H_{n+1}$ in (\ref{inti}) for
the I-hierarchy):
\begin{equation}\label{rpm}
A_n^- h_n = \gamma\,h_{n+1} A^-_n,\qquad  h_n A_n^+= \gamma\,A^+_n h_{n+1}
\end{equation}
where $\gamma$ is for a sign.
As a consequence of \eqref{rpm}, the operators $A^\pm_n$ will also
intertwine $h_n^2$ and $h_{n+1}^2$:
\begin{equation}\label{rpm2}
A_n^- h_n^2 = h_{n+1}^2 A^-_n,\qquad  h_n^2 A_n^+= A^+_n h_{n+1}^2
\end{equation}
where $A_n^{\pm}={(A_n^{\mp})}^\dag$. There are two solutions to this problem  (up
to a multiplicative constant):

\noindent a) True intertwining operators ($\gamma=1$)
\begin{equation}\label{rx}
R^-_n = \left(\begin{array}{cc}
 a^-_n & i(\mu_{n+1} - \mu_n)\\[1.ex]
0 &  a^-_{n+1}
\end{array}\right) ,\qquad R_n^- h_n = h_{n+1} R^-_n
\end{equation}

\noindent b) Anti-intertwining operators ($\gamma=-1$)
\begin{equation}\label{rxx}
T^-_n = \left(\begin{array}{cc}
 a^-_n & -i(\mu_{n+1} + \mu_n)\\[1.ex]
0 & - a^-_{n+1}
\end{array}\right) ,\qquad T_n^- h_n = - h_{n+1} T^-_n
\end{equation}
One couple, $R_n^\pm$, plays the role of the usual intertwining operators;
but there is a new pair $T_n^\pm$ carrying a minus sign that we call anti-intertwining. This possibility is specific of Dirac-like Hamiltonians where the spectrum includes a negative sector, as we will see in the next subsection.
\subsection{Spectrum of the Dirac hierarchy $\boldsymbol{ h_n}$}

We will compute the eigenfunctions and eigenvalues of each Dirac
Hamiltonian (\ref{hn1}),
\begin{equation}\label{hne}
h_n \Psi_n^k = \varepsilon_n^k \Psi_n^k
\end{equation}
where $\varepsilon_n^k$ designs the $k$th excited level of the energy eigenvalues of Hamiltonian $h_n$ and $\Psi_n^k$ is a corresponding spinor eigenfunction.
According to the square of $h_n$ given in (\ref{sqr}), we will express the spectrum and eigen-spinors  of $h_n$ ($\varepsilon_n^k$ and $\Psi_n^k$) in terms of the eigenfunctions of the scalar Schr\"odinger Hamiltonians $H_n$  ($\mu_{n+k}^2$, $\psi_n^k$) and $H_{n+1}$ ($\mu_{n+k+1}^2$, $\psi_{n+1}^k$), respectively. We make the natural ansatz
\begin{equation}\label{diracS}
\Psi_n^{k}= \left(\begin{array}{c}
\alpha\,\psi_n^k
\\[1.5ex]
i\beta\,\psi_{n+1}^{k-1}
\end{array}\right),\qquad \alpha,\beta\in \mathbb C
\end{equation}
where $\alpha,\beta$ must be computed by substituting in the
eigenvalue equation for $h_n$ defined in (\ref{hn1}). There are two
signs for the energy. Then, to specify the energy sign $\pm$ we will write
$\varepsilon_n^{k\pm}$ and $ \Psi_n^{k\pm}$ for the respective
eigenfunctions. After some computations we get the following
solutions (they are not normalized).
\begin{itemize}
\item Positive spectrum
\begin{equation}\label{ps}
\begin{array}{ll}
\varepsilon_n^{k+} = \mu_{n+k},\qquad
&\Psi_n^{k+}= \left(\begin{array}{c}
\sqrt{\mu_{n+k}+\mu_n}\,\psi_n^k
\\[1.5ex]
-i\sqrt{\mu_{n+k}-\mu_n}\,\psi_{n+1}^{k-1}
\end{array}\right)\ ,\qquad k=1,2,\dots
\\[4.ex]
{\rm Ground\ level}:
\\[1.ex]
\varepsilon_n^{0+} = \mu_n,\quad
&\Psi_n^{0+}= \left(\begin{array}{c}
\psi_n^0
\\[1.ex]
0
\end{array}\right)\ ,\qquad k=0
\end{array}
\end{equation}

\item Negative spectrum
\begin{equation}\label{ns}
\varepsilon_n^{k-} = -\mu_{n+k},
\qquad
\Psi_n^{k-}= \left(\begin{array}{c}
\sqrt{\mu_{n+k}-\mu_n}\,\psi_n^k
\\[1.5ex]
i\sqrt{\mu_{n+k}+\mu_n}\,\psi_{n+1}^{k-1}
\end{array}\right)\ , \qquad k=1,2,\dots
\end{equation}
\end{itemize}
Notice some properties of the spectrum.

\begin{enumerate}
\item
The positive and negative solutions $\Psi_n^{k\pm}$ are solutions
of the square of $h_n$ with the same positive eigenvalue:
\begin{equation}\label{hsquare}
(h_n)^2 \Psi_n^{k\pm} =
\left(\begin{array}{cc}
H_n & 0\\[1.ex]
0 & H_{n+1}
\end{array}\right)\left(\begin{array}{c}
\sqrt{\mu_{n+k}\pm\mu_n}\,\psi_n^k
\\[1.5ex]
\mp i\sqrt{\mu_{n+k}\mp\mu_n}\,\psi_{n+1}^{k-1}
\end{array}\right) =
\mu_{n+k}^2 \,\Psi_n^{k\pm}
\end{equation}

\item The ground level $k=0$ exists only
for positive energies, while the ``ground negative'' energy level
is reached for $k=1$.

\item
The eigenfunctions for positive energy have
the upper component with bigger norm than the lower one, while eigenfunctions
of negative energy have oposite character: the norm of the lower component is
bigger than the upper.

\item
Eigenfunctions of positive and negative energies are orthogonal.

\end{enumerate}

\subsection{The spectrum for a null factorization energy (Dirac--Weyl like equation)}

Up to now we have assumed that each Dirac Hamiltonian $h_n$, as it is defined in (\ref{hn1}), has the factorization energy (it may play the role of mass) given by $\mu_n$. However, we could be interested in the spectrum
of a particular $h_{n_0}$ with $n=n_0$,  having a different value of the factorizacion energy, for instance 
$\mu_{n_0}=0$, (it would correspond to a null mass case means that the Dirac--Weyl like equation).

In principle, the Schr\"odinger Hamiltonian hierarchy (increasing (\ref{ih}) or decreasing (\ref{dh})) will have $\mu_{n_0}^2\neq 0$.
Then, we will subtract
from all the Hamiltonian hierarchy $\{ H_n= a_n^+a_n^-+\mu_n^2\}$ the same value: $\mu_{n_0}^2 = E_{n_0}^0$. Hence,
we will have the new sequence $\{ \tilde H_n= a_n^+a_n^-+ \tilde \mu_n^2\}$
where
\[
\tilde \mu_n^2 =  \mu_n^2 -  \mu_{n_0}^2
\]
This new hierarchy has the same properties, intertwining, eigenfunctions, etc. except the spectrum that now will be displaced the same amount:
\[
\tilde E_n^k = \tilde E^0_{n+k} = \tilde \mu_{n+k}^2= E^0_{n+k}- E^0_{n_0},
\qquad \tilde E^0_{n_0}= 0
\]
In other words, the ground energy of the displaced Hamiltonian $\tilde H_{n_0}$ is zero and the corresponding
Dirac Hamiltonian $\tilde h_{n_0}$ will have zero mass:
\begin{equation}
\tilde h_n = \left(\begin{array}{cc}
\tilde \mu_n & i a^+_n\\[1.ex]
-i a^-_n & -\tilde\mu_n
\end{array}\right)\,,\qquad
\tilde h_{n_0} = \left(\begin{array}{cc}
0 & i a^+_{n_0}\\[1.ex]
-i a^-_{n_0} & 0
\end{array}\right)\,,\qquad \tilde\mu_{n_0} = 0
\end{equation}

The formulas for the positive (\ref{ps}) or negative (\ref{ns}) spectrum
$\tilde \varepsilon_n^{k\pm}$ and wave functions $\tilde \Psi_n^{k\pm}$
will still be valid, with the replacement of the new
values of $\tilde\mu_{n}$ in their respective expressions.

\subsection{Eigenfunctions annihilated by the intertwining operators}

Next, we want to know what kind of eigenfunctions annihilate the intertwining
operators $R_n^-$ and $T_n^-$. Since they are first order linear $2\times 2$-matrix differential operators, they must annihilate two linearly independent spinor functions.
\begin{itemize}
\item
Eigenfunctions annihilated by $R_n^-$

1) The ground state of positive energies:
\[
\Psi_n^{0+}= \left(\begin{array}{c}
\psi_n^0
\\[1.ex]
0
\end{array}\right)\,,
\qquad \varepsilon_n^{0+}= \mu_n
\]

2) The ``ground state'' of negative energies:
\[
\Psi_n^{1-}= \left(\begin{array}{c}
\sqrt{\mu_{n+1}-\mu_n}\,\psi_n^1
\\[1.5ex]
i\sqrt{\mu_{n+1}+\mu_n}\,\psi_{n+1}^{0}
\end{array}\right)\,,
\qquad \varepsilon_n^{1-}= - \mu_{n+1}
\]

\item
Eigenfunctions annihilated by $T_n^-$

1) The ground state of positive energies:
\[
\Psi_n^{0+}= \left(\begin{array}{c}
\psi_n^0
\\[1.ex]
0
\end{array}\right)\,,
\qquad \varepsilon_n^{0+}= \mu_n
\]

2) The first excited state of positive energies:
\[
\Psi_n^{1+}= \left(\begin{array}{c}
\sqrt{\mu_{n+1}+\mu_n}\,\psi_n^1
\\[1.5ex]
-i\sqrt{\mu_{n+1}-\mu_n}\,\psi_{n+1}^{0}
\end{array}\right)\,,
\qquad \varepsilon_n^{1+}=  \mu_{n+1}
\]
\end{itemize}
In conclusion, the matrix intertwining operators $R_n^-$ and $T_n^-$ annihilate
 some of lowest energy states of the Hamiltonians $h_n$, in a similar way as
the factor operators $a_n^-$ do for the ground states of $H_n$. However, in the
matrix case there are more options due to the two sectors (positive and negative) of the spectrum and
the fact that each intertwining operator annihilates two independent states.

\subsection{Symmetries from intertwining operators}

As we have seen in a precedent subsection, the spectrum of this type
of Dirac Hamiltonians $h_n$ have simple degeneracy: the eigenspaces are
one-dimensional. This means that there are no additional independent symmetry,
or in other words, any symmetry must be a function of the Hamiltonian $h_n$.

A method to build symmetries is by means of the product of two intertwining operators in opposite direction; for instance $S_n$ obtained by the product $R_n^+R_n^-$
will be a symmetry of $h_n$:
\[
R_n^-h_n = h_{n+1}R_n^-, \qquad    h_n R_n^+ = R_n^+h_{n+1} \\, \implies\,
[S_n,h_n] = 0
\]
Therefore, this symmetry must be a function of $h_n$. After a computation
we find:
\[
S_n =R_n^+R_n^- = 
(h_n - \mu_n)(h_n+\mu_{n+1})
\]
and the commutator is
\[
R_n^+R_n^--R_{n-1}^-R_{n-1}^+=h_n(\mu_{n+1}-2\mu_n+\mu_{n-1}) +\mu_n(\mu_{n-1}-\mu_{n+1})
\]
 This result seems reasonable since we have seen before that the operator
 $R_n^-$ annihilates eigenvectors with eigenvalues $\mu_n$ and
 $-\mu_{n+1}$ and the same happens with the operators on the right hand side of the last equality.

 The symmetry obtained from the other intertwining (anti-intertwining) operator:
 \[
 S'_n = T_n^+T_n^-
 \]
 will have a similar expression:
 \[
 S'_n= (h_n - \mu_n)(h_n-\mu_{n+1})
 \]
and in this case the commutator has the form
\[
T_n^+T_n^--T_{n-1}^-T_{n-1}^+=-h_n(\mu_{n+1}+2\mu_n+\mu_{n-1}) -\mu_n(\mu_{n-1}-\mu_{n+1})
\]
 This is consistent with the eigenvectors (and their eigenvalues)
 annihilated by $T_n^-$ (see the previous subsection).
\subsection
{Intertwining operators for $m_0$-mass Dirac-like Hamiltonians}
\medskip
In this subsection, we will mention briefly how our approach to Dirac-like
matrix Hamiltonians is compatible with a constant mass $m_0$.

Firstly, we define the $4\times4$ Dirac-like
Hamiltonian ${\cal H}_n$ with mass $m_0$ from the previous $h_n$ as follows \cite{sandro,alonso,demir21}:
\begin{equation}\label{hhn}
{\cal H}_n = \left(\begin{array}{cc}
m_0 c^2 &h_n\\[1.5ex]
h_n & -m_0 c^2
\end{array}\right)
\end{equation}
Then, we can also  introduce $4\times4$ intertwining operators for the hierarchy
${\cal H}_n$ , in terms of the previous $2\times 2$ operators, $R_n^\pm=(R_n^\mp)^{\dag}$ and $T_n^\pm=(T_n^\mp)^{\dag}$. In fact, we have two extensions for each pair having the following form (together with their adjoint operators and taking $c=1$):
\begin{equation}\label{rrn}
{\cal R}_n^-= \left(\begin{array}{cc}
R_n^- &0\\[1.5ex]
0& R_n^-
\end{array}\right),\qquad
{\tilde {\cal R}}_n^-= \left(\begin{array}{cc}
-m_0 M^-&R_n^- \\[1.5ex]
R_n^- & m_0 M^-
\end{array}\right)
\end{equation}

\begin{equation}\label{ttn}
{\cal T}_n^-= \left(\begin{array}{cc}
-T_n^- &0\\[1.5ex]
0& T_n^-
\end{array}\right),\qquad
{\tilde {\cal T}}_n^-= \left(\begin{array}{cc}
m_0 M^-&-T_n^- \\[1.5ex]
T_n^- & m_0 M^-
\end{array}\right)
\end{equation}
where 
\[
M^-:=
-i\sigma^+=-i(\sigma_1+i\sigma_2)
\] 
is a constant matrix which satisfies the following (anti)commutation relations with the $2\times2$ Hamiltonians $h_n$:
\[
M^- h_n+h_{n+1}M^-= -2R_n^-,\qquad M^- h_n-h_{n+1}M^-=-2 T_n^-\]
It can be checked  that all these  global intertwining operators fulfil
the usual type of intertwining relations, for example:
\[
{\cal R}_n^-{\cal H}_n={\cal H}_{n+1}{\cal R}_n^-,
\qquad
{\cal T}_n^-{\cal H}_n={\cal H}_{n+1}{\cal T}_n^-
\]
The adjoint operators ${\cal R}_n^+=({\cal R}_n^-)^\dagger$ and ${\cal T}_n^+=({\cal T}_n^-)^\dagger$
will satisfy the opposite intertwining relations.
Finally, we can say that in fact we have obtained a four dimensional vector space of global intertwining operators for the same hierarchy of $4\times 4$ Hamiltonians
${\cal H}_{n}$. In principle, this is surprising because in scalar Hamiltonian hierarchies there is only one set of factor intertwining operators for the same hierarchy. In our case, the degeneracy of the positive and negative energy levels of each Hamiltonian ${\cal H}_n$ allows that the intertwining operators be non unique.

Let us obtain the eigenfunctions and the spectrum of ${\cal H}_n$. The eigenvalue equation is
\begin{equation}\label{hep}
{\cal H}_n\, \Xi_n^{k s} ={\cal E}^k_{n \pm}\, \Xi_n^{k s}
\end{equation}
where the energy is ${\cal E}^k_{n \pm}$ and each state $\Xi_n^{k s}$ is composed of two eigen-spinors of $h_n$ of same type:
\begin{equation}\label{hepf}
\Xi_n^{k s}=
\left(\begin{array}{c}
\Psi_n^{k s}
\\[1.5ex]
\Phi_n^{k s}
\end{array}\right)\,
\end{equation}
For instance, $\Psi_n^{k s}$ corresponds to the eigenvalue $\varepsilon_n^{k s}$ of the $n$-Hamiltonian $h_n$, $k$-energy level, and energy sign $s=\pm$.
Using ${\cal H}_n$ in the eigenvalue equation, we get two coupled equations:
\begin{equation}\label{pf}
m_0 c^2 \Psi _n^{k s}+ h_n \Phi_n^{k s}= {\cal E}^k_{n \pm}\Psi_n^{k s},   
\qquad  -m_0 c^2 \Phi_n^{k s} + h_n \Psi_n^{k s}= {\cal E}^k_{n \pm} \Phi_n^{k s}\,
\end{equation}
From these equations, we obtain the following  relations
\begin{equation}\label{pf1}
\Phi_{n\pm}^{k s}=\frac{h_n \Psi_n^{k s}}{{\cal E}^k_{n\pm}+m_0 c^2}, 
\qquad \Psi_{n\pm}^{k s}=\frac{h_n \Phi_n^{k s}}{{\cal E}^k_{n\pm}-m_0 c^2}\,
\end{equation}
Substituting these in  (\ref{pf}), we get the equations for the spinors $\Psi$ and $\Phi$
\begin{equation}\label{pf2}
h_n^2 \Psi_n^{k s}=({({\cal E}_n^k)^2-m_0^2c^4}) \Psi_n^{k s}, 
\qquad h_n^2 \Phi_n^{k s}=({({\cal E}_n^k)^2-m_0^2c^4}) \Phi_n^{k s}
\end{equation}
taking into account $h_n^2 \Psi_n^{k s}=(\varepsilon_n^{k s})^2 \Psi_n^{k s}$, we get the eigenvalues 
\begin{equation}\label{espm0}
{\cal E}^k_{n\pm}=\pm \sqrt{(\varepsilon_n^{k s})^2+m_0^2c^4}= \pm \sqrt{\mu_{n+k}^2+m_0^2c^4}
\end{equation}
 where $E_n^k=(\varepsilon_n^k)^2=\mu_{n+k}^2$, as shown in (\ref{hsquare}). Then, we find the eigenfunctions:
\begin{equation}\label{hepf1}
\Xi_{n +}^{k s}=
\left(\begin{array}{c}
\Psi_{n }^{k s}
\\[1.5ex]
\Phi_{n +}^{k s}
\end{array}\right)\,,
\qquad
\Phi_{n +}^{k s}=\frac{s\,\mu_{n+k} \Psi_n^{k s}}{\sqrt{\mu_{n+k}^2+m_0^2c^4}+m_0 c^2}
\end{equation}
or 
\begin{equation}\label{hepf2}
\Xi_{n -}^{k s}=
\left(\begin{array}{c}
\Psi_{n{-}}^{k s}
\\[1.5ex]
\Phi_{n}^{k s}
\end{array}\right)\,,
\qquad
\Psi_{n-}^{k s}=-\frac{s\,\mu_{n+k} \Phi_n^{k s}}{\sqrt{\mu_{n+k}^2+m_0^2c^4}+m_0 c^2}
\end{equation}
We notice that the spectrum (\ref{espm0}) for a non-vanishing $m_0$ is
symmetric in the positive and negative eigenvalue sectors, contrary to $h_n$.
The eigenvalues have double degeneracy $s=\pm$, except for the ground energy
of each sector which have simple degeneracy. 

\section
{Example: the Trigonometric P\"oschl--Teller Potential}

An example of a hierarchy of increasing type is the following sequence of PT Schr\"odinger Hamiltonians
\begin{equation}\label{initial}
H_n(x) = -\partial_{x x}
+\frac{(n+1/2)(n-1/2)}{\sin^2 x}  ,\qquad x\in (0,\pi),
\qquad n\in \mathbb N
\end{equation}
$H_n(x)$ satisfy the factorization given by (\ref{ih}). The factor
operators and factor energies are
\begin{equation}\label{facener}
a_n^\pm(x) = \mp\partial_x - (n+1/2) \cot x,\qquad
\mu_n^2 = (n+1/2)^2,\qquad n=0,1,2,\dots
\end{equation}
Therefore, the associated Dirac Hamiltonians $h_n$ are defined, according to (\ref{hn1}), by
\begin{equation}\label{ptn}
h_n = \left(\begin{array}{cc}
n+1/2 & i (-\partial_x - (n+1/2) \cot x)\\[1.5ex]
-i (+\partial_x - (n+1/2) \cot x) & -n-1/2
\end{array}\right)
\end{equation}
Their square reproduce the Hamiltonians $H_n$ and $H_{n+1}$, as it is given in (\ref{sqr}).
The explicit intertwining operators, given by (\ref{rx}) and (\ref{rxx}), take here the expressions
\begin{equation}\label{rxb}
R^-_n = \left(\begin{array}{cc}
  \partial_x - (n+1/2) \cot x & i\\[1.5ex]
0 &  \partial_x - (n+3/2) \cot x
\end{array}\right)
\end{equation}
and
\begin{equation}\label{txc}
T^-_n = \left(\begin{array}{cc}
 \partial_x - (n+1/2) \cot x & -i(2n+2)\\[1.5ex]
0 & - (\partial_x - (n+3/2) \cot x)
\end{array}\right)
\end{equation}
In the next subsection, we will explain the origin of the scalar and matrix hierarchies together with the (anti)intertwining operators.


\subsection
{Dirac-like equations for systems on the ${\cal S}^2$ sphere}
\medskip

Here we want to show how these two Hamiltonians, the scalar PT Schr\"odinger one
(\ref{initial}) and the matrix PT Dirac (\ref{ptn}), can be obtained by reduction of a scalar and a spinor
particle systems respectively, defined on the ${\cal S}^2$ sphere. We will start with the scalar case,
which is more or less well known, and then we will introduce the one of Dirac type, which is new up to our knowledge.
\medskip

\noindent
{\bf\em (a) Scalar system}
\medskip

\noindent
Let us consider a system on the two-dimensional sphere ${\cal S}^2$
defined by the points $(x,y,z)\in \mathbb R^3$ satisfying
\[
x^2+y^2+z^2 =1
\]
The $SO(3)$ group will act in the usual way on the scalar wavefunctions $\tilde\psi({\bf x})$ defined on
${\cal S}^2$, 
\begin{equation}\label{so3}
U(g)\tilde \psi({\bf x})= \tilde \psi(g^{-1}{\bf x}),\qquad g\in SO(3)
\end{equation}
If we use Cartesian $(x,y,z)$ or spherical $(r,\theta,\phi)$ coordinates 
\[
x = r \sin\theta \cos\phi,\qquad
y = r \sin\theta \sin \phi,\qquad
z = r \cos\theta
\]
the Hermitian generators $L_i$, $i=1,2,3$ (or $L_x,L_y,L_z$) of $SO(3)$ represented by the (\ref{so3}) given by:
\begin{equation}\label{genso3}
\begin{array}{l}
L_x= -i(y\partial_z-z\partial_y) = i(\sin\phi\, \partial_\theta +\cos\phi \cot\theta\,\partial_\phi)
\\[2.ex]
L_y = -i(z\partial_x - x\partial_z)= i(-\cos\phi\, \partial_\theta + \sin \phi\cot\theta\, \partial_\phi)
\\[2.ex]
L_z= -i(x\partial_y - y \partial_x) = i(-\partial_\phi)
\end{array}
\end{equation}
We recall the notation of the  lowering and raising angular momentum operators:
\begin{equation}\label{dif1}
L_\pm =L_x\pm iL_y,\qquad L_\pm = e^{\pm i\phi}(\pm \partial_\theta
+ i \cot \theta \partial_\phi)
\end{equation}
such that
\[
[L_z,L^\pm ] = \pm L^\pm ,\qquad [L^-,L^+] = -2 L_z
\]

The quadratic Casimir operator $\tilde H:= {\cal C} = L_x^2 +L_y^2 + L_z^2$ plays
the role of a free two-dimensional Hamiltonian in the space of  wavefunctions defined on the sphere,  with eigenvalues $\ell(\ell+1)$ corresponding to the eigenfunctions of each irreducible representation:
\begin{equation}\label{htilde}
 \tilde H \tilde\psi_\ell = \left(-\partial_{\theta\theta} - \cot\theta \partial_\theta
-\frac{\partial_{\phi\phi}}{\sin^2\theta}\right) \tilde\psi_\ell =
\ell(\ell+1)\tilde\psi_\ell,
\qquad \ell=0,1,\dots 
\end{equation}
We will eliminate the term with the first partial derivative by
means of the transformation
\begin{equation}\label{trans}
\tilde \psi = \frac1{\sqrt{\sin\theta}}\,\psi
\quad \implies\quad
\partial_\theta \ \to \ \partial_\theta-\frac12 \cot\theta;
\quad \tilde H\  \to\  H-1/4
\end{equation}
After applying this transformation, we get the following Hamiltonian equation (with $H:=\tilde H +1/4$) in a explicitly Hermitian equivalent form
\begin{equation}\label{scalar}
  H \psi_\ell = \left(-\partial_{\theta\theta}
-\frac{1/4+\partial_{\phi\phi}}{\sin^2\theta}\right) \psi_\ell =
(\ell+1/2)^2\,\psi_\ell,
\qquad \ell=0,1,\dots
\end{equation}
In summary, the above Schr\"odinger 
Hamiltonian can be identified, in the realization (\ref{trans}),
with the $SO(3)$ Casimir operator (plus the constant $1/4$)
\begin{equation}\label{hscalar}
H = {\boldsymbol L}^2 +1/4
\end{equation}
If we take separated solutions, which are also eigenfunctions of the  generator $L_z$ defined in (\ref{genso3}),
\begin{equation}\label{seppsi}
\psi_{\ell n}(\theta,\phi) = \psi_{\ell n}(\theta)e^{in\phi},\qquad
L_z\psi_{\ell n}(\theta,\phi)= n\, \psi_{\ell n}(\phi,\theta),\qquad
|n|\leq \ell
\end{equation}
the component $\psi_{\ell n}(\theta)$ will satisfy the reduced equation
\begin{equation}\label{hhn}
  H_n \psi_{\ell n} = \left(-\partial_{\theta\theta}
+\frac{(n+1/2)(n-1/2)}{\sin^2\theta}\right) \psi_{\ell n} =
(\ell+1/2)^2\,\psi_{\ell n},
\quad n= 0,\pm1,\dots \pm\ell
\end{equation}
If we fix the eigenvalue by means of the parameter $\ell$ then, the Hamiltonians $H_n$ having this
eigenvalue are those parametrised by $n = 0,\pm 1,\dots ,\pm \ell$.
The states span the $2\ell+1$--dimensional representation $``\ell"$ of $SO(3)$ with eigenvalue
$\varepsilon_{\ell n}= (\ell+1/2)^2$.

On the other hand, once selected a value $n$, the spectrum of the reduced Hamiltonian $H_n$ is given by the eigenvalues
\[
E_n^k = (n+k+1/2)^2,\qquad  k=0,1,\dots,\quad n+k:=\ell
\]
Equation (\ref{hhn}) is the same as the initial one (\ref{initial}) of
the PT potential (with the variable $\theta$ instead of $x$) and the factorization energies $\mu_n^2$ coincide
with $E_n^0$. Therefore, we have the following connection between the eigenvalue notation
$(E_n^k, \psi_n^k(\theta))$ and the notation of $SO(3)$ representations ($\varepsilon_{\ell n} 
= (\ell+1/2)^2, \psi_{\ell n})$:
\[
\left\{\begin{array}{ll}
 H_n\psi_n^k = E_n^k\psi_n^k ,\
 &E_n^k= (n+k+1/2)^2
 \\[2.ex]
H_n\psi_{\ell n}  = \varepsilon_{\ell n}\psi_{\ell n} ,
\ &\varepsilon_{\ell n}= (\ell+1/2)^2
\end{array}\right.
 \!\!\implies\ \ n+k=\ell\ \implies \left\{
\begin{array}{ll}
\psi_n^k = \psi_{(n+k)n}\  
\\[2.ex]
\psi_{\ell n} = \psi_n^{\ell-n}  
\end{array}\right.
\]
This relationship tell us that the eigenfunctions of this hierarchy $H_n$
are essentially spherical harmonics (subjected to the above transformation):
\begin{equation}\label{spha}
\psi_n^k(\theta) e^{i n \phi}\propto  {\sqrt{\sin \theta}}\,Y_{\ell
n}(\theta,\phi),\quad \ell = n+k
\end{equation}
In particular, for $k=0$ have the form:
\[ \psi_n^0(\theta) e^{i n \phi} \propto  {\sqrt{\sin
\theta}}\,Y_{nn}(\theta,\phi)\] 
If we apply the transformation
(\ref{trans}) to the orbital angular operators (\ref{dif1}), they
become
\begin{equation}\label{dif2}
L_+ = e^{i\phi}\Big(\partial_\theta +  \cot \theta(-1/2+i \partial_\phi)\Big),
\quad
L_- = \Big(-\partial_\theta +
\cot \theta(-1/2+ i\partial_\phi)\Big)e^{-i\phi}
\end{equation}
where now it is explicit the adjoint relation $(L_+)^\dagger= L_-$. If we allow them to act on
the separated solutions (\ref{seppsi}) we find the reduced expressions
\begin{equation}\label{dif3}
L^+_n =  \partial_\theta -  \cot \theta(1/2+n),
\quad
L^-_n = -\partial_\theta -   \cot \theta(1/2+n)
\end{equation}
which coincide with the intertwining operators $a^-_n$ (corresponds to $L^+_n$) and $a^+_n$
(to $L^-_n$), given in (\ref{facener}).

Next, we want to carry out a similar construction of the Dirac-like
Hamiltonian (\ref{ptn}) by means of a spinor realization of $SO(3)$ (or better, its universal covering $SU(2)$) on the sphere ${\cal S}^2$.

\bigskip

\noindent
{\bf\em (b) Spinor system}
\medskip

\noindent
We will consider an spinor system defined on the unit sphere by means of two-component
wavefunctions $\Psi({\bf x})$, ${\bf x}\in {\cal S}^2$. The action of the group $SU(2)$ on a function of this space has
the usual form
\[
U(g)\Psi({\bf x})= D(g) \Psi(g^{-1}{\bf x}),\qquad g\in SU(2)
\]
where $D$ is for the spin $1/2$ fundamental representation of $SU(2)$, and the action of $g\in SU(2)$ on
${\cal S}^2$ is through the homomorphism of $SU(2)$ onto $SO(3)$.
The generators corresponding to this action are given by the total rotations (orbital plus spin):
\[
J_i = L_i(\theta,\phi) + \frac12 \sigma_i:=L_i(\theta,\phi) +
S_i,\qquad i=1,2,3
\]
with $\sigma_i$  ($\sigma_1=\sigma_x, \sigma_2=\sigma_y,
\sigma_3=\sigma_z$) being the three $2\times2$ Pauli spin matrices. These
operators close the $su(2)\approx so(3)$ Lie algebra. In this case,
the Casimir operator is
\begin{equation}\label{cas}
{\cal C}_{\rm s} = {\boldsymbol J}^2 =
{\boldsymbol L}^2 + 2{\boldsymbol L}{\boldsymbol S}+ {\boldsymbol S}^2
\end{equation}
It is clear that ${\boldsymbol L}^2$ and ${\boldsymbol S}^2$ are also
Casimir operators for the spinor system (both commute with all the
generators $J_i$). Therefore, another Casimir operator commuting with the operators ${\boldsymbol J}$ is
\begin{equation}\label{hdsu2}
\tilde h:= {\cal C}_{\rm so} = 2{\boldsymbol L}{\boldsymbol S}= 2(\sum_k L_kS_k)
\end{equation}
which has the form of a spin-orbit coupling term.
Notice that this is a first order Casimir in the momentum operators
and it has a matrix character due to $\boldsymbol S$; by these reasons we will identify this operator as a Dirac-like Hamiltonian $\tilde h$.

In order to simplify some calculations, we will make use of the  lowering and raising spin operators:
\[
S_\pm = S_x\pm iS_y,\qquad S_+ = \left(\begin{array}{ll}
0 & 1\\[1.ex]
0 & 0 \end{array}\right)\,, \qquad S_- = \left(\begin{array}{ll}
0 & 0\\[1.ex]
1 & 0 \end{array}\right)
\]
Then, we have
another useful expression of $\tilde h$ defined by (\ref{hdsu2}) in terms of $L_\pm$ (given in (\ref{dif2})) and $S_\pm$:
\begin{equation}\label{hpm}
\tilde h=   L_-S_+ + L_+ S_- + 2L_zS_z
\end{equation}
After replacement of the operators in (\ref{hpm}), the final explicit expression for
$\tilde h$ is
\begin{equation}\label{diracex}
\tilde h 
  =
\left(\begin{array}{cc}
-i \partial_\phi & \Big(-\partial_\theta +
\cot \theta(-1/2+ i\partial_\phi)\Big)e^{-i\phi}
\\[2.ex]
e^{i\phi}\Big(\partial_\theta +  \cot \theta(-1/2+i \partial_\phi)\Big)
& i\partial_\phi
\end{array}\right)
\end{equation}
Since $\tilde h$ commutes with the generators ${\boldsymbol J}$, the eigenfunctions
$\Psi_{jm}$ can be labeled by the parameters $j$ and $m$ corresponding to the eigenvalues of ${\boldsymbol J}^2$ and $J_z$:
\begin{equation}\label{jmn}
{\boldsymbol J}^2\Psi_{jm}= j(j+1) \Psi_{jm},\qquad
J_z\Psi_{jm} = m\Psi_{jm}
\end{equation}
If we apply the matrix Hamiltonian (\ref{diracex}) on the eigenfunctions
of the generator $J_z$, with undetermined value of ${\boldsymbol J}^2$,
\begin{equation}\label{mn}
J_z\Psi_{m} = m \Psi_{m}  \ \implies \ \Psi_{m}(\theta,\phi) =
\left(\begin{array}{l} \varphi_{m}(\theta)e^{i(m-1/2)\phi}
\\[1.ex]
\chi_{m}(\theta)e^{i(m+1/2)\phi}
\end{array}\right)
\end{equation}
then, we get the reduced Hamiltonian in the $\theta$ variable
\begin{equation}\label{hm}
\tilde h_m = \left(\begin{array}{cc}
m-1/2 &  -(\partial_\theta +m
\cot \theta)
\\[2.ex]
 -(-\partial_\theta +m \cot \theta)
& -m-1/2
\end{array}\right)
\end{equation}
which is similar to $h_n$ proposed at the beginning of this section
in (\ref{ptn}); in fact
\begin{equation}\label{hmn}
\tilde h_m +1/2 = h_n, \quad 
\quad m=n+1/2
\end{equation}
We must have in mind that, according to its definition $m$ is a pure
half-integer, i.e. $m-1/2= n$ and $m+1/2=n+1$, are integer values of the orbital angular momentum $L_z$ corresponding
to the spinor components in (\ref{mn}).
We will also use the notation of (\ref{hmn}) for non reduced Hamiltonians: $\tilde h +1/2 = h$.
\bigskip

We can also compute the square of $\tilde h$ with the aim of recovering the Schr\"odinger equation
(\ref{scalar}):
\begin{equation}\label{square}
(2{\boldsymbol L}{\boldsymbol S})^2 = {\boldsymbol L}^2 - 2{\boldsymbol L}{\boldsymbol S}\
\implies
\
\Big(2{\boldsymbol L}{\boldsymbol S} +\frac12\Big)^2 = {\boldsymbol L}^2 +\frac14
:= {\boldsymbol H}
\end{equation}
In other words,
we must
choose $h=\tilde h +1/2$, where $\tilde h$ is defined in (\ref{diracex}) and square it as in (\ref{square}) to get 
\begin{equation}\label{square2}
h^2 = {\boldsymbol H}
\end{equation}
Therefore, by taking the square (in this modified way) of the Dirac operator $h$, we get the scalar Hamiltonian (\ref{scalar}),
just in a similar manner as from the Dirac equation we can get the Klein-Gordon equation.

In particular, if we apply this formula to the  Hamiltonian
$\tilde h_m$ in (\ref{hm}), with $m=n+1/2$, we get the scalar reduced Hamiltonians
$H_n$ and $H_{n+1}$ in a diagonal matrix, as it was shown in (\ref{hhn}),
\begin{equation}\label{square3}
\Big(\tilde h_m+\frac12\Big)^2 =\left(\begin{array}{cc}
H_n & 0
\\[1.ex]
0  & H_{n+1}
\end{array}\right)\,, \qquad m=n+1/2
\end{equation}
We are always making use of the relation $n= m-1/2$ and $n+1= m+1/2$ for the spinor components.

In this way, we have explained the origin of the first order
Dirac-like matrix Hamiltonians $h_n$ corresponding to the  scalar
Hamiltonians $H_n$ from the factorization method \cite{infeld}. Next, we will determine the intertwining operators of the
sequence $h_n$ by means of the symmetries of $h$.
\bigskip

\noindent
{\bf\em (c) Symmetries and ``anti-symmetries''}

\begin{itemize}
\item[1)] Symmetries $J_\pm, J_z$.

In the following we will compute the symmetries corresponding to the
lowering and raising operators $J_\pm$ for the eigenfunctions
$\Psi_{jm}$ of $J_z$ and ${\boldsymbol J}^2$. Taking into account the expressions of
(\ref{dif2}), we get
\[
J_+=L_++S_+=\left(\begin{array}{cc}
e^{i\phi}\Big(\partial_\theta +  \cot \theta(-1/2+i \partial_\phi)\Big)
& 1
\\[1.5ex]
0 & e^{i\phi}\Big(\partial_\theta +  \cot \theta(-1/2+i \partial_\phi)\Big)
\end{array}\right)
\]

\[
J_-=L_-+S_-=\left(\begin{array}{cc}
\Big(-\partial_\theta +
\cot \theta(-1/2+ i\partial_\phi)\Big)e^{-i\phi}
& 0
\\[1.5ex]
1 & \Big(-\partial_\theta +
\cot \theta(-1/2+ i\partial_\phi)\Big)e^{-i\phi}
\end{array}\right)
\]
When $J_+$ act on $\Psi_{jm}$ and $J_-$ on $\Psi_{jm+1}$  their effect on the components $\varphi_m(\theta)$
and $\chi_m(\theta)$ of (\ref{mn}) turns into
\[
J^+_m=\left(\begin{array}{cc}
\partial_\theta -m  \cot \theta
& 1
\\[1.5ex]
0 &  \partial_\theta -(m+1) \cot \theta
\end{array}\right)\,;
\ \
J^-_m=\left(\begin{array}{cc}
-\partial_\theta -m  \cot \theta
& 0
\\[1.5ex]
1 &  -\partial_\theta -(m+1) \cot \theta
\end{array}\right)
\]
These expressions are related with the previous ones shown in (\ref{rxb}) by means of a simple equivalence
($R^\pm_n\approx J^\mp_m$, $m=n+1/2$). We should also include the symmetry $J_z = L_z + S_z$ which is used
to define the basis of eigenvectors $\Psi_{jm}$ (\ref{jmn}).
\medskip

\item[2)] Anti-symmetries $T^\pm, T_3$.

We will also  find the anti-symmetries $T^\pm$ which anti-commute with
the Hamiltonian $h$ and behave like shift operators of $J_z$:
\[
T^\pm  (\tilde h +1/2) = - (\tilde h+1/2) \, T^\pm,\qquad \quad
[J_z,T^\pm] = \pm T^\pm
\]
Due to their anti-commutation with $\tilde h +1/2$, the operators
$T^\pm$ change a positive energy state of $\tilde h +1/2$ into
another one with the same absolute but opposite sign energy. It can
be checked that such  anti-symmetries are given by
\[
T^+ = L_+S_3 - L_3 S_+,\qquad (T^+)^\dagger = T^-= L_-S_3 - L_3 S_-
\]
where $S_3 =S_z$ and  $L_3 =L_z$. When they act on the basis $\Psi_{jm}$, $T^\pm$ acquire an equivalent expression to that given in (\ref{txc}).
Besides the above anti-symmetry operators, we must include a third one,
called $T_3$ defined by:
\[
T_3 = -\frac12 (L_+S_- - L_-S_+)
\]
We can also check that the set of anti-symmetries $T_\pm,T_3$ constitute a vector operator under the action of
${\boldsymbol J}$; in other words, they satisfy the commutation rules
\[
[J_+,T^+] = 0, \quad [J_+,T^-]= 2 T_3, \quad [J_\pm,T_3] = \mp T^\pm,
\quad [J_3,T^\pm] = \pm T^\pm,\quad [J_3,T_3] =0
\]

The operators $T^\pm, T_3$ are ``odd" operators in the sense that
connect the spaces of positive and negative energies states. The
pair $T^\pm$ besides changing the energy sign, they also change the
$J_3=J_z$ eigenvalue $m$, but $T_3$ keeps $m$ invariant. All of them
change the eigenvalue $j$ of ${\boldsymbol J}^2$ to $j-1$ or to
$j+1$ if the initial space was of positive or negative energies,
respectively. This is consistent with the following commutation
rule,
\[
[{\boldsymbol J}^2,T^+] = 2 (\tilde h+1/2) T^+
\]
\end{itemize}

The  existence of  anti-symmetry operators does not imply that the
positive and negative eigenspaces are completely symmetric, as we will see later.  Notice that this type of nontrivial anti-symmetry operators
 is a remarkable property of this kind of  Dirac-like Hamiltonians.
\bigskip

\noindent
{\bf\em (d) Spectrum and eigenfunctions}
\medskip

In this subsection, we will compute the discrete spectrum and
eigenfunctions in the same way as we have seen above. Later on we  will
identify the eigenfunctions in relation with the symmetries.

First of all we choose the eigenfunctions with a definite value of
the momentum $J_z$ given by (\ref{mn}).
As we have seen in (\ref{square3}), the square of $\tilde h_m+1/2$
is a diagonal matrix with $H_n$ and $H_{n+1}$ in the diagonal. These
two scalar Hamiltonians have the same eigenvalues
$E_n^k=(n+k+1/2)^2$ for the respective eigenfunctions $\psi_n^k$ and
$\psi_{n+1}^{k-1}$. Therefore, this suggest the following form of
the spinor eigenfunctions of $\tilde h_m+1/2$:
\[
\Psi_n^k=
\left(\begin{array}{c}
\alpha\, \psi_n^k(\theta) e^{i (m-1/2)\phi}
\\[1.5ex]
\beta\, \psi_{n+1}^{k-1}(\theta) e^{i (m+1/2)\phi}
\end{array}\right) ,\qquad m=n+1/2,\quad k=0,1,2\dots
\]
where the constants $\alpha,\beta\in \mathbb C$ must be found from the
eigenvalue equation:
\[
(\tilde h_m +1/2)\Psi_n^k = \tilde \varepsilon\,  \Psi_n^k ,\qquad m=n+1/2
\]
We have positive energy solutions
\[
\tilde \varepsilon_n^{k+} = + (n+k + 1/2),\quad
\Psi_n^{k+}=
\left(\begin{array}{c}
\alpha_n^{k+}\, \psi_n^k(\theta) e^{i (m-1/2)\phi}
\\[1.5ex]
\beta_n^{k+}\, \psi_{n+1}^{k-1}(\theta) e^{i (m+1/2)\phi}
\end{array}\right) ,\qquad
m=n+1/2,\quad k=0,1,2\dots
\]
and negative energy solutions
\[
\tilde \varepsilon_n^{k-} = - (n+k + 1/2),\quad
\Psi_n^{k+}=
\left(\begin{array}{c}
\alpha_n^{k-}\, \psi_n^k(\theta) e^{i (m-1/2)\phi}
\\[1.5ex]
\beta_n^{k-}\, \psi_{n+1}^{k-1}(\theta) e^{i (m+1/2)\phi}
\end{array}\right) ,\qquad
m=n+1/2,\quad k=1,2\dots
\]
with eigenfunctions determined by the coefficients
\[
 \alpha_n^{k+}=\sqrt{2n+k+1}, \ \beta_n^{k+}=-\sqrt{k}, \qquad
 \alpha_n^{k-}=\sqrt{k}, \ \beta_n^{k-}=  \sqrt{2n+k+1}
\]
The zero ground energy level is positive:
\[
\tilde \varepsilon_n^{0+} = + (n + 1/2)
\]
and its wave function is
\[
\Psi_n^{0+}(\theta,\phi)=
\left(\begin{array}{c}
 \psi_n^0(\theta) e^{i (m-1/2)\phi}
\\[1.5ex]
0
\end{array}\right)
\]
Next, we will interpret the eigenfunctions $\Psi_n^{k\pm}$ and their spinor
components in terms of the commuting symmetries of $\tilde h$: $J_z$, ${\boldsymbol L}^2$ and ${\boldsymbol J}^2$.

\begin{itemize}
\item[$J_z$:]

The label $n$ of the eigenfunctions $\Psi_n^{k\pm}$ means that, in fact
they are eigenfunctions of $J_z$ with eigenvalue $m=n+1/2$:
\[
J_z\Psi_n^{k\pm}= m \Psi_n^{k\pm},  \qquad m= n+1/2
\]

\item[${\boldsymbol L}^2$:] We know from the previous scalar case that the
scalar functions $\psi_n^k(\theta) e^{i (m-1/2)}$ and
$\psi_{n+1}^{k-1}(\theta) e^{i (m+1/2)}$ (which are de components of
the spinors $\Psi_n^{k\pm}$) are eigenfunctions of ${\boldsymbol
L}^2$ with eigenvalue ``$\ell(\ell+1)$''.

Therefore, the spinor eigenfunctions of $\tilde h$ are also eigenfunctions
of ${\boldsymbol L}^2$ with the same eigenvalue
\[
({\boldsymbol L}^2+1/4)\Psi_n^{k\pm}= (\ell+1/2)^2 \Psi_n^{k\pm},
\qquad\ell= n+k, 
\quad m= n+1/2
\]

\item[${\boldsymbol J}^2$:]
The above eigenfunctions of $\tilde h$ are also eigenfunctions of ${\boldsymbol J}^2$.
We will make use of expression (\ref{cas}) in order to see this:
\[
{\boldsymbol J}^2\Psi_n^{k+}=
\big( {\boldsymbol L}^2 + \tilde h +{\boldsymbol S}^2\big) \Psi_n^{k+}=
( \ell(\ell+1)+\ell +3/4)\Psi_n^{k+}= (\ell+1/2)(\ell+3/2)\Psi_n^{k+}
\]
In other words, it can be written as
\[
{\boldsymbol J}^2\Psi_n^{k+}=
 (j^+)(j^++1)\Psi_n^{k+},\qquad j^+=\ell+1/2
\]
In the same way, we check that
\[
{\boldsymbol J}^2\Psi_n^{k-}=
 (j^-)(j^-+1)\Psi_n^{k+},\qquad j^-=\ell-1/2
\]
We could explain these results from the fact that the spinor wave functions $\Psi$
belong to tensor products of the representations ``$\ell$'' from the orbital part and $s=1/2$ from spin: $\Psi\in \ell\otimes 1/2$. As we know (from the
composition of angular momenta):
\[
\ell\otimes 1/2 = j^+\oplus j^-,\qquad j^\pm = \ell\pm 1/2
\]
Then, the positive energy spinors $\Psi_n^{k+}$ belong to the
representation of the total angular momentum $ j^+$, while the
negative spinors $\Psi_n^{k-}$ to $ j^-$.
\end{itemize}

In summary, the eigenfunctions of $\tilde h+1/2$ could be labeled in two equivalent ways with the notations $\Psi_n^{k\pm}$ and $\Psi_{j^\pm\ell m}$:
\[
\begin{array}{lll}
\Psi_n^{k+}= \Psi_{j^+\ell m};\
&\tilde \varepsilon_n^{k+} = \ell+1/2,
\qquad &\ell= n+k,\quad m=n+1/2,\quad j^+= \ell+1/2
\\[2.ex]
\Psi_n^{k-}= \Psi_{j^-\ell m}; \ &\tilde \varepsilon_n^{k-} = -(\ell+1/2),
\qquad &\ell= n+k,\quad m=n+1/2,\quad j^-= \ell-1/2
\end{array}
\]

As a consequence, this result confirms us that the degeneracy of
positive energy $\tilde \varepsilon_n^{k+}$ levels is greater than
the corresponding negative level $\tilde \varepsilon_n^{k-}$. This
is due to the fact that the positive eigenspace is the support of
the representation $j^+ =\ell+1/2$, whose dimension is $2\ell+2$,
while the negative eigenspace supports $j^-$ of dimension $2\ell$.

\section{Hierarchies of 1D Dirac Hamiltonians: Decreasing Case}

Next, we will carry out this program for a decreasing Hamiltonian hierarchy $H_n$ characterized by the
factorization
\begin{equation}\label{hd}
H_{n -1}=-\partial_{xx} + V_n(x) = a^+_{n -1} a^-_{n -1} - \mu_{n -1}^2= a^-_{n} a^+_{n} - \mu_{n}^2,\qquad
n=1,2,\dots
\end{equation}
which was given in \eqref{dh}. Then, we can form a Dirac-like Hamiltonian hierarchy
$h_n$ defined by
\begin{equation}\label{hhyp}
h_n = \left(\begin{array}{cc}
\mu_n & i a^-_n\\[1.ex]
i a^+_n & -\mu_n
\end{array}\right)
\end{equation}
The square of $h_n$ gives a pair of Schr\"odinger Hamiltonians of the D-hierarchy, with an extra minus sign
\begin{equation}\label{sqr2}
h_n^2 = - \left(\begin{array}{cc}
H_{n-1} & 0\\[1.ex]
0 & H_{n}
\end{array}\right) := -{\boldsymbol H}_n
\end{equation}
Thus, we see that the Dirac-like Hamiltonians (\ref{hhyp}) in this case have a relevant difference with the increasing hierarchy: they are non-Hermitian. In fact, we could say that they are $\sigma_3$-Hermitian \cite{mosta02,oscar18} in the sense that
\begin{equation}\label{noherm}
h_n = \sigma_3\, h_n^\dagger\, \sigma_3:=h_n^\sharp
\end{equation}
However, since $\sigma_3$ commutes with $h_n^2$ given in (\ref{sqr2}) 
we find that this is
Hermitian: $(h_n^2)^\dagger= h_n^2$.

Next, we are looking for a pair of intertwining operators $A^\pm_n$ between the matrix Hamiltonians $h_n$ and $h_{n-1}$ in the same way as (\ref{inti2}) for the scalar D-hierarchy:
\begin{equation}\label{rp}
A_n^- h_n = \gamma h_{n-1} A^-_n,\qquad  
h_n A_n^+= \gamma A^+_n h_{n-1}
\end{equation}
where $A^\pm_n= \sigma_3 (A_n^\mp)^\dagger \sigma_3$ and $\gamma$ is for a sign.
As a consequence, the operators $A^\pm_n$ will intertwine
$h_n^2$ and $h_{n-1}^2$:
\begin{equation}\label{rpm2}
A_n^- h_n^2 = h_{n-1}^2 A^-_n,\qquad  h_n^2 A_n^+= A^+_n h_{n-1}^2
\end{equation}
The solution for the case of intertwining operators ($\gamma=1$) is 
\begin{equation}\label{rrx}
R^-_n = \left(\begin{array}{cc}
 a^-_{n-1} &  0\\[1.ex]
i(\mu_{n-1} - \mu_n) &  a^-_{n}
\end{array}\right) ,\qquad R_n^- h_n = h_{n-1} R^-_n
\end{equation}
While the anti-intertwining operators ($\gamma=-1$) are $T^\pm_n$:
\begin{equation}\label{ttx}
T^-_n = \left(\begin{array}{cc}
 a^-_{n-1} & 0\\[1.ex]
i(\mu_{n-1} + \mu_n) & -a^-_{n}
\end{array}\right) ,\qquad T_n^- h_n = - h_{n-1} T^-_n
\end{equation}

\subsection{Spectrum of the Dirac hierarchy for decreasing case}
The spectrum and spinor eigenfunctions of the Hamiltonian $h_n$ (\ref{hhyp})  can also be found following the same procedure as before. We get the solutions:
\begin{itemize}
\item Positive spectrum
\begin{equation}
\varepsilon_n^{k+} = \mu_{n-k},\quad
\Psi_n^{k+}
=
\left(\begin{array}{c} \sqrt{\mu_n+\mu_{n-k}}\, \psi_{n-1}^{k-1}
\\[1.5ex]
i\sqrt{\mu_n-\mu_{n-k}}\,\psi_n^k
\end{array}\right),\qquad k=1,2,\dots
\end{equation}

\item Negative spectrum
\begin{equation}
\begin{array}{ll}
\varepsilon_n^{k-} = -\mu_{n-k}, \quad &\Psi_n^{k-}=
\left(\begin{array}{c} \sqrt{\mu_n-\mu_{n-k}}\,\psi_{n-1}^{k-1}
\\[1.5ex]
i\sqrt{\mu_n+\mu_{n-k}}\,\psi_n^k
\end{array}\right),\qquad k=1,2,\dots
\\[4.ex]
{\rm Ground\ level}:
\\[1.ex]
\varepsilon_n^{0-} = -\mu_{n},\quad &\Psi_n^{0-}=
\left(\begin{array}{c} 0
\\[1.ex]
\psi_n^0
\end{array}\right),\qquad k=0
\end{array}
\end{equation}
\end{itemize}

Remark that in this Dirac-like version of the scalar D-hierarchies
the ``absolute'' ground state has negative energy. Now, the role of antiparticle
states are played by the positive energy spectrum. This is an example of
non-Hermitian
matrix Hamiltonian, as it is explicit in (\ref{hhyp}), where the discrete energy levels are real and the eigenfunctions square integrable.

\subsection{Eigenfunctions annihilated by the intertwining operators}

Next, we want to know what kind of eigenfunctions annihilate the
intertwining operators $R_n^-$ and $T_n^-$.
\begin{itemize}
\item
Eigenfunctions annihilated by $R_n^-$

1) The ground state of negative energies:
\[
\Psi_n^{0-}=
\left(\begin{array}{c} 0
\\[1.ex]
\psi_n^0
\end{array}\right)\,,
\qquad \varepsilon_n^{0-}= -\mu_n
\]

2) The ``ground state'' of positive energies:
\[
\Psi_n^{1+}
=
\left(\begin{array}{c} \sqrt{\mu_n+\mu_{n-1}}\, \psi_{n-1}^{0}
\\[1.5ex]
i\sqrt{\mu_n-\mu_{n-1}}\,\psi_n^1
\end{array}\right)\,,
\qquad \varepsilon_n^{1+}= \mu_{n-1}
\]

\item
Eigenfunctions annihilated by $T_n^-$

1) The ground state of negative energies:
\[
\Psi_n^{0-}=
\left(\begin{array}{c} 0
\\[1.ex]
\psi_n^0
\end{array}\right)\,, \qquad \varepsilon_n^{0-}= -\mu_n
\]

2) The first excited state of negative energies:
\[
\Psi_n^{1-}=
\left(\begin{array}{c} \sqrt{\mu_n-\mu_{n-1}}\,\psi_{n-1}^{0}
\\[1.5ex]
i\sqrt{\mu_n+\mu_{n-1}}\,\psi_n^1
\end{array}\right)\,,
\qquad \varepsilon_n^{1-}= -\mu_{n-1}
\]

\end{itemize}
\subsection
{Example: The hyperbolic P\"oschl--Teller potential}

The following sequence of hyperbolic PT Schr\"odinger Hamiltonians
\begin{equation}\label{pt}
H_n(x) = -\partial_{xx} - \frac{(n+1/2)(n-1/2)}{\cosh^2x},\qquad x\in
(-\infty,\infty),\qquad n\in \mathbb N
\end{equation}
are of decreasing type and satisfy the factorization (\ref{dh}). The
factor operators and factor energies are
\begin{equation}\label{a11mp}
a_n^\pm(x) = \mp\partial_x + (n-1/2)\tanh x,\qquad \mu_n^2 =
(n-1/2)^2,\qquad n=1,2,\dots
\end{equation}
Therefore, the associated Dirac Hamiltonians $h_n$ are defined by
\begin{equation}\label{mpt}
h_n = \left(\begin{array}{cc}
 (n-1/2) & i (\partial_x + (n-1/2) \tanh x)\\[1.ex]
i (-\partial_x + (n-1/2) \tanh x) & -(n-1/2)
\end{array}\right)
\end{equation}
and the intertwining operators by
\begin{equation}\label{rxb2}
R^+_n = \left(\begin{array}{cc}
 -\partial_x + (n-3/2) \tanh x & -i\\[1.ex]
0 &  -\partial_x + (n-1/2) \tanh x
\end{array}\right)
\end{equation}
\begin{equation}\label{rxc2}
T^+_n = \left(\begin{array}{cc}
-\partial_x + (n-3/2) \tanh x & i\,(2n-2)\\[1.ex]
0 & - (-\partial_x + (n-1/2) \tanh x)
\end{array}\right)
\end{equation}
\subsection
{Dirac-like equations for systems on the ${\cal H}^2$ hyperboloid}

We can get the hyperbolic PT scalar Hamiltonian (\ref{pt}) and the matrix PT Hamiltonian corresponding to
(\ref{mpt}) from separation of variables of a system defined on a hyperboloid in a similar way as
that employed in the
trigonometric case in the space of functions on the sphere. In order to
get the potential in (\ref{pt}) we start with one sheet hyperboloid
(this choice is different from the two sheets hyperboloid and it has some peculiarities).
We have made this election in order to work with representations of eigenfunctions in the
angular $\phi$ variable and to obtain the specific form of the hyperbolic potential
(\ref{pt}).
\medskip

\noindent
{\bf\em (a) The scalar system}
\medskip

\noindent
Consider the hyperbolic surface ${\cal H}^2$,
\begin{equation}\label{hyp}
x^2+y^2-z^2=1
\end{equation}
parametrised with pseudo-spherical coordinates:
\begin{equation}\label{pseu}
x=\cosh \chi \cos\phi,\qquad y =\cosh \chi \sin \phi,\qquad z=\sinh
\chi
\end{equation}
The surface (\ref{hyp}) is invariant under the group $SO(2,1)$ and the scalar wavefunctions defined on ${\cal H}^2$ constitute the support space of a quasi-regular representation $U$,
\begin{equation}\label{hypU}
U(g)\tilde\psi({\bf x}) = \tilde \psi(g^{-1}{\bf x}),\qquad g\in SO(2,1)
\end{equation}
The generators of the group $SO(2,1)$ are given by: one of them, $L_z$, generating rotations around the $z$-axis and two, $L_x$ and $L_y$, of hyperbolic transformations around the $x$ and $y$ axes. A differential realization in  Cartesian coordinates is
\[
L_x= y \partial_z + z\partial_y,\qquad L_y = - x\partial_z -
z\partial_x,\qquad L_z= -x\partial_y + y\partial_x
\]
If we pass to the pseudo-spherical coordinates (\ref{pseu}) and multiply by the imaginary unit $i$, they take the form
\begin{equation}\label{lpm11}
L_x = i(\sin \phi \,\partial_\chi + \tanh \chi \cos \phi\,
\partial_\phi),\quad
L_y = i(-\cos \phi \,\partial_\chi + \tanh \chi
\sin \phi\,\partial_\phi),\quad
L_z = i(-\partial_\phi)
\end{equation}
Their commutation rules close the $so(2,1)\approx su(1,1)$ Lie algebra
\[
[L_z, L_x] = i\,L_y,\qquad
[L_z,L_y] = - i\,L_x,\qquad
[L_x,L_y]=-i\,L_z
\]
We build the lowering and raising operators from the realization (\ref{lpm11}),
\begin{equation}\label{lpm11b}
L_{\pm} = L_x \pm i L_y = e^{\pm i \phi}
\Big(\pm \partial_\chi + i\,\tanh \chi \partial_\phi\big)
\end{equation}
In this basis the $so(2,1)\approx su(1,1)$ Lie algebra is
\[
[L_z,L_\pm] = \pm L_\pm,\qquad [L_-,L_+] = 2L_z
\]
The Casimir operator, which we will identify with the scalar Schr\"odinger Hamiltonian, has the form
\begin{equation}\label{cas11}
{\cal C}:= \tilde H =  L_x^2 + L_y^2 - L_z^2 = L_+L_- -L_z(L_z-1)
\end{equation}
The discrete series of  unitary irreducible representations (uir's) of $SO(2,1)$ are characterized by the Casimir eigenvalues
\cite{perelomov,vilenkin,bargmann}
\begin{equation}\label{h11}
\tilde H \tilde\psi = -\lambda(\lambda-1)\tilde \psi, \qquad \lambda=1, 2, 3,\dots
\end{equation}
In terms of the pseudo-spherical coordinates ${\cal C}$ ($=\tilde H$) takes the expression
\[
\tilde H  = -\partial_{\chi\chi} - \tanh \chi\,\partial_\chi
+\frac{\partial_{\phi\phi}}{\cosh^2\chi}
\]
There is a first order term in $\partial_\chi$ that can be eliminated by means of the following
transformation:
\begin{equation}\label{transf}
\tilde \psi = \frac1{\sqrt{\cosh \chi}} \psi \ \implies \
\partial_\chi \to \partial_\chi -\frac12\,\tanh \chi
\end{equation}
Then, the displaced Hamiltonian $H:=\tilde H -\frac14$ becomes
\begin{equation}\label{newh}
H = \tilde H -\frac14 = -\partial_{\chi\chi}
+\frac{1/4 +\partial_{\phi\phi}}{\cosh^2\chi}
\end{equation}
The eigenvalues of this Hamiltonian corresponding to the uir's of $SO(2,1)$
are, according to (\ref{h11}),
\[
 H \psi = -(\lambda-1/2)^2 \psi, \qquad \lambda=1, 2,\dots
\]
If we consider a basis in this representation of common eigenfunctions of ${\cal C}$ and $L_z$,
\[
\psi(\chi,\phi) = \psi_n(\chi)e^{in\phi},\qquad n\in {\mathbb N}
\]
we find
\begin{equation}\label{ptm}
 H_n \psi_n = \left(-\partial_{\chi\chi}
-\frac{(n+1/2)(n-1/2)}{\cosh^2\chi}\right) \psi_n=-(\lambda-1/2)^2 \psi_n, \qquad \lambda=1,2,3,\dots,n
\end{equation}
where $n\geq \lambda\geq 1$. Therefore, once fixed $n$ then the number of bound states is finite: $k=0,1,\dots, n-1$. In case we fixed the eigenvalue $\lambda$, the potentials that may support such a negative level  are labeled by $n\geq \lambda$.

We will use the notation $\psi_n^k$ for the eigenfunction corresponding to:
\[
\psi_n^k=\left\{\begin{array}{ll}
{\rm Potential}\quad &V_n = -\frac{(n+1/2)(n-1/2)}{\cosh^2\chi}
\\[2.ex]
{\rm Energy \ level} \quad &E_n^k = -(n-k-1/2)^2,\qquad k= 0,1\dots n-1
\\[2.ex]
{\rm Representation} &\lambda= n-k,\qquad L_z\psi_n^k=n \psi_n^k,\qquad \lambda= n,n-1,\dots, 1
\end{array}\right.
\]
We can check that indeed this equation coincides with the initial hyperbolic PT
equation (\ref{pt}) 
as well as the eigenvalues (\ref{a11mp}).

On the other hand, the lowering and raising operators, when they act between the Hamiltonians $H_n$ and $H_{n-1}$, after the transformation (\ref{transf}),
become
\begin{equation}\label{lpm11c}
\begin{array}{l}
L_{+} = L_x + i L_y =
\big(+\partial_\chi -(n-1/2)\tanh \chi  \big)
\\[2.ex]
L_{-} = L_x - i L_y =
\big(- \partial_\chi -(n-1/2)\tanh \chi  \big)
\end{array}
\end{equation}
These operators coincide with the intertwining operators $a_n^\pm$ given before in (\ref{a11mp}).
\medskip

\noindent
{\bf\em (b) The spinor system}
\medskip

\noindent
Next, we will consider the spinor representation of $so(2,1)\approx su(1,1)$
defined in the space of spinor wavefunctions $\Psi({\bf x})$ on the hyperboloid (\ref{hyp}).
In this space, the generators of $SU(1,1)$, which will be called $K_i$, are defined by
\begin{equation}\label{ks}
K_i = L_i+ S_i,\qquad i=1,2,3
\end{equation}
where the orbital generators ($L_1=L_x, L_2=L_y, L_3=L_z$) have been given in (\ref{lpm11}), while the spinor generators
($S_1=S_x, S_2=S_y, S_3=S_z$) have been chosen as
\[
\textstyle
S_x= i\frac12\sigma_x,\qquad
S_y= i\frac12\sigma_y,\qquad
S_z= \frac12\sigma_z
\]
and therefore, $S_\pm = S_x\pm i S_y$ are
\begin{equation}\label{spmhyp}
S_+ = \left(\begin{array}{cc}
0 & i\\[1.ex]
0 & 0\end{array}\right)\,,\qquad
S_- = \left(\begin{array}{cc}
0 & 0\\[1.ex]
i & 0\end{array}\right)
\end{equation}
The matrices $S_x,S_y$ are anti-Hermitian, while $S_z$ is Hermitian. They  satisfy the $su(1,1)$ Lie algebra:
\[
[S_z,S_x] = i\, S_y,\qquad
[S_z,S_y] = -i\, S_x,\qquad
[S_x,S_y] = -i\, S_z
\]
or
\[
[S_z,S_\pm] = \pm S_\pm,\qquad [S_-,S_+] = 2 S_z
\]
Then, the Casimir operator of this realization is
\begin{equation}\label{caspin11}
{\cal C}_s = {\boldsymbol K}^2 = (L_x+S_x)^2 + (L_y+S_y)^2- (L_z+S_z)^2=
{\boldsymbol L}^2 +2{\boldsymbol L}{\boldsymbol S} +{\boldsymbol S}^2 = -\nu(\nu-1)
\end{equation}
where the cross terms constitute the matrix Hamiltonian $h$,
\begin{equation}\label{hspin21}
h = 2{\boldsymbol L}{\boldsymbol S} = 2(L_xS_x+L_yS_y - L_zS_z) = L_+S_- + L_-S_+ -2L_zS_z
\end{equation}
and ${\boldsymbol S}^2= -3/4$. We will use the spinor expressions (\ref{spmhyp}) and the
orbital operators $L_\pm$ of (\ref{lpm11b}) after the transformation (\ref{transf}),
\[
L_\pm = e^{\pm i \phi} \big( \pm \partial_\chi + (i\partial_\phi \mp  1/2) \tanh \chi \big)
\]
to be replaced in the expression (\ref{hspin21}) of $h$ in order to find the explicit matrix Hamiltonian
\begin{equation}\label{hspinh}
h = \left(\begin{array}{cc}
i\partial_\phi & i e^{-i\phi}(-\partial_\chi  +(i\partial_\phi +1/2)\tanh\chi)
\\[2.ex]
i(+\partial_\chi +(i\partial_\phi +1/2)\tanh\chi)e^{i\phi} & -i \partial_\phi
\end{array}\right)
\end{equation}

As we know from the beginning this Hamiltonian is not explicitly Hermitian since the
matrix generators $S_x$ and $S_y$ were not Hermitian. Let us write the form of $h$ when
it acts on simultaneous eigenfunctions $\Psi_{\nu m}$ of the Casimir ${\cal C}$ (\ref{caspin11}) and $K_z$:
\begin{equation}\label{mn2}
K_z\Psi_{\nu m} = m \Psi_{\nu m}  \ \implies \
\Psi_{\nu m}(\chi,\phi) = \left(\begin{array}{l}
\varphi_{m}(\chi)e^{i(m-1/2)\phi}
\\[1.ex]
\xi_{m}(\chi)e^{i(m+1/2)\phi}
\end{array}\right)
\end{equation}
Then, the reduced Hamiltonian in the variable $\chi$, using this basis, is
\begin{equation}\label{hspinh2}
h_m = \left(\begin{array}{cc}
-m+1/2 & i  (-\partial_\chi  -m\tanh\chi)
\\[2.ex]
i  (\partial_\chi  -m\tanh\chi) & m+1/2
\end{array}\right)
\end{equation}

\noindent
The next step is to compute the square of $h$:
\begin{equation}
h^2 = -{\boldsymbol L}^2 +h
\end{equation}
According to (\ref{newh}) $H={\boldsymbol L}^2-1/4$ so that finally we get
\begin{equation}\label{squarehyp}
(h-1/2)^2 = -H =
-\left(-\partial_{\chi\chi} +\frac{1/4 +\partial_{\phi\phi}}{\cosh^2\chi}\right)
\end{equation}
If we take the square $(h_m-1/2)^2$ of the reduced Hamiltonian (\ref{hspinh2}), we
find
\begin{equation}\label{sqred}
(h_m-1/2)^2 = - \left(\begin{array}{cc}
H_{n-1} & 0
\\[1.ex]
0 & H_{n}\end{array}\right)
\end{equation}
where
$H_m $ coincides with $H_n$ of (\ref{ptm}), provided $m=n-1/2$.
Let us make some remarks.
\medskip

\noindent
(a)
These two formulas for the square of $h$ are reasonable since the square of a real spectrum
of $h$ will always produce a positive spectrum, that is, this square must have oposite sign
than the scalar Hamiltonians $H$, which have a negative spectrum as shown in (\ref{ptm}).
\medskip

\noindent
(b)
The matrix hyperbolic Hamiltonian $h$ given in (\ref{hspinh}), or its reduced form
(\ref{hspinh2}), are not explicitly Hermitian, therefore this seems an inconsistency.
However, $h$ although non-Hermitian it has a real discrete spectrum with square integrable solutions. We will show this property in the following subsection by computing  the eigenvalues and eigenvectors.
\medskip

\noindent
(c)
Let us notice that, as a consequence of the previous remark, this spinor representation of $SU(1,1)$ is not unitary. However, there is a Hermitian
invariant product given by
\[
\langle \Phi,\Psi\rangle = \int
\Phi^\dagger(\chi,\phi) \sigma_3 \Psi(\chi,\phi)\, {\rm d}\chi {\rm d}\phi
\]
where the invariant measure on the hyperboloid is hidden in the wavefunctions
by the transformation (\ref{transf}).  One can check that indeed, the reduced Hamiltonian (\ref{hspinh}) is Hermitian with respect to $\sigma_3$: $h^\dagger \sigma_3
= \sigma_3 h$. This is a non positive definite product which is consistent
with the non-unitary character.

\bigskip

\noindent
{\bf\em (c) Spectrum and eigenfunctions}
\medskip

If we have in mind the diagonal form of the square $(h-1/2)^2$, and at the same time
we write the operators in the entries of the matrix $h_m$ given in (\ref{hspinh2}) we
guess that the eigenfunctions will have the following form, making use of  the  factor operators $a_m^\pm$ of (\ref{a11mp}):
\begin{equation}\label{eigenhyp}
\left(\begin{array}{cc}
-m  & -i a_m^-
\\[1.5ex]
-i a_m^+ &  m
\end{array}\right)
\left(\begin{array}{l}
\alpha \psi_{m-1}^{k-1}
\\[1.5ex]
i\,\beta \psi_{m}^{k}
\end{array}\right)
=
\varepsilon_n^k
\left(\begin{array}{l}
\alpha \psi_{m-1}^{k-1}
\\[1.5ex]
i\,\beta \psi_{m}^{k}
\end{array}\right)
\end{equation}
Taking into account that the action of these factor operators is
\[
a_m^-\psi_{m}^{k}= \sqrt{\mu_m^2 - \mu_{m-k}^2}\  \psi_{m-1}^{k-1},\qquad
a_m^+\psi_{m-1}^{k-1}= \sqrt{\mu_m^2 - \mu_{m-k}^2}\  \psi_{m}^{k},\qquad
\mu_{m-k}^2= (m-k)^2
\]
Then, we get the eigenvalues of $h$:
\begin{equation}\label{espAH}
\varepsilon_n^{k\pm} =\pm (m-k)\,, \qquad m=n-1/2
\end{equation}
The corresponding eigenspinors $\Psi_n^{k\pm}$ are given by their coefficients $\alpha^\pm$ and $\beta^\pm$:
\[
\alpha^- = \sqrt{k}\,,\ \ \beta^- = \sqrt{2m-k}\,,\qquad
\alpha^+ = \sqrt{2m-k}\,,\ \ \beta^+ =   \sqrt{k}
\]
The ground state of these eigenfunctions is negative:
\[
\varepsilon_n^{0-} =-m,\qquad \Psi_n^{0-}=
\left(\begin{array}{l}
0
\\[1.5ex]
\psi_{m}^{0}
\end{array}\right)
\]
We notice that the greater spinor component of the negative spectrum eigenfunctions is the lower one with spin $-1/2$. However, the greater component of the positive spectrum spinor eigenfunctions is
the upper corresponding to spin $1/2$.
The degeneracy is infinite for each positive or negative eigenvalue. In the following we will determine
the $SU(1,1)$ representations corresponding to the eigenvalues of the same absolute value but opposite sign:
$\varepsilon_n^{k\pm}=\pm (n-k-1/2)$.
In order to find these representations, we will characterize the eigenfunctions $\Psi_n^{k\pm}$ and their spinor components in terms of the commuting symmetries of $h$: $K_z$, ${\boldsymbol L}^2$ and ${\boldsymbol K}^2$.

\begin{itemize}
\item[$K_z$:]

The label $n$ of the spinor functions $\Psi_n^{k\pm}$ means that they are eigenfunctions of $K_z$ with eigenvalue $m=n-1/2$:
\[
K_z\Psi_n^{k\pm}= m \Psi_n^{k\pm},  \qquad m= n-1/2
\]

\item[${\boldsymbol L}^2$:] We know that the
scalar functions $\psi_n^k(\theta) e^{i (m+1/2)\phi}$ and
$\psi_{n-1}^{k-1}(\theta) e^{i (m-1/2)\phi}$ (which are de components of the spinors $\Psi_n^{k\pm}$)
are eigenfunctions of ${\boldsymbol L}^2$ with eigenvalue $\lambda=n-k$.
Therefore, the spinor eigenfunctions of $ h$ are also eigenfunctions
of ${\boldsymbol L}^2$ with the same eigenvalue
\[
({\boldsymbol L}^2-1/4)\Psi_n^{k\pm}= -(\lambda-1/2)^2 \Psi_n^{k\pm},
\qquad \lambda= n-k, 
\qquad m= n-1/2
\]

\item[${\boldsymbol K}^2$:]
The above eigenfunctions of $ h$ are also eigenfunctions of ${\boldsymbol K}^2$.
We will make use of expression (\ref{cas}) in order to see this:
\[
{\boldsymbol K}^2\Psi_n^{k-}=
\big( {\boldsymbol L}^2 + \tilde h +{\boldsymbol S}^2\big) \Psi_n^{k-}=
(-\lambda(\lambda-1)-\lambda +1/4)\Psi_n^{k-}= -(\lambda+1/2)(\lambda-1/2)\Psi_n^{k-}
\]
In other words, we can write as
\[
{\boldsymbol K}^2\Psi_n^{k-}=
 (\nu^+)(\nu^+-1)\Psi_n^{k-},\qquad \nu^+=\lambda+1/2
\]
and in a similar way
\[
{\boldsymbol K}^2\Psi_n^{k+}=
 (\nu^-)(\nu^- -1)\Psi_n^{k+},\qquad \nu^-=\lambda-1/2
\]
We could explain these results from the fact that the spinor wave functions $\Psi$
belong to tensor products of the representations $\lambda$ from the orbital part and $s=1/2$ from spin: $\Psi\in \lambda\otimes 1/2$. As we know from the
composition of  ``angular momenta'':
\[
\lambda\otimes 1/2 = \nu^+\oplus \nu^-,\qquad \nu^\pm = \lambda\pm 1/2
\]
Then, the negative energy spinors $\Psi_n^{k-}$ belong to the representation of the total ``angular momentum'' $\nu^+$, while the
positive spinors $\Psi_n^{k+}$ to $ \nu^-$.

\end{itemize}

In conclusion, we have obtained a real spectrum for a non-Hermitian matrix Hamiltonian.
\bigskip

\noindent
{\bf\em (d) Symmetries and ``anti-symmetries''}

\begin{itemize}
\item[1)] Symmetries $K_\pm, K_3$.

In the same way as for $su(2)$ the spinor symmetries in the case of $su(1,1)$
are given by
\[
K_+= L_++S_+,\qquad K_-= L_-+S_-,\qquad K_3= L_3+S_3
\]
Their restriction to the eigenfunctions of $K_3=K_z$ give rise to the
intertwining operators $R^\pm_n$.

\item[2)] Anti-symmetries

We will also  find the anti-symmetries $T^\pm$ which anti-commute with
the Hamiltonian $h_n-1/2$,
\[
T^\pm  (\tilde h -1/2) = - (\tilde h-1/2) \, T^\pm,\qquad
\quad [K_3,T^\pm] = \pm T^\pm
\]
and have the same form as in the trigonometric case:
\[
T^+ = L_+S_3 - L_3 S_+,\qquad  T^-= L_-S_3 - L_3 S_-,\qquad
T_3 = \frac12(L_+S_- - L_-S_+)
\]

\end{itemize}

The spectrum of Dirac-Weyl like equations, symmetries in terms of intertwining operators and global intertwining operators for $4\times 4$ Dirac-like Hamiltonians can also be obtained following the same procedure as above. However,  in order to shorten the length of the paper we have not discussed them here. 

\section{Conclusions and remarks}

We have introduced a kind of matrix  Dirac-like Hamiltonians which are in close
correspondence with factorizable Schr\"odinger Hamiltonians. The scalar intertwining 
operators, $a^\pm_n$ and the factorization energies $\pm\mu_n^2$, are used as ingredients to construct the matrix Hamiltonian and the intertwining matrix
operators. This type of Dirac Hamiltonians have appeared in many previous references
taking part in different problems. 
Here, we have tried to explain them by a reduction
process of spinor systems defined on curved spaces. 
We have called $h_n$ (the reduced cases in one variable) or 
$h$ (in two coordinates of the surface) to these Dirac-like Hamiltonians. In the case
of spherical symmetry they are called Dirac operators \cite{marten58}.

We have worked out two simple examples.
One of them defined on a sphere ${\cal S}^2$, giving rise to a matrix version
of a trigonometric P\"oschl-Teller potential. The symmetries and anti-symmetries were
characterized as well as the degeneracy of the energy levels. In particular the anti-symmetries
interchanged the sign of energy. The second example  which was defined in a two dimensional hyperboloid
${\cal H}^2$ gave rise to a matrix non-Hermitian Hamiltonian with a real spectrum, which is related to a scalar hyperbolic P\"oschl-Teller potential. 
The scalar systems were obtained as the square of the Dirac-like
systems, just in a similar way as the Dirac and Klein-Gordon equations are related. 
We expect to complete this picture in
the near future by finding the Dirac-like cases of all the scalar factorizable potentials
with or without the mass $m_0$.
In particular they will need higher space dimensions and higher gamma matrices. 
Another point that we pursue is to identify the role of this kind of Dirac-like Hamiltonians; 
it is reasonable that they be related with the symmetries of different classes of Dirac systems,
since our building relies only on symmetry considerations. We hope that this kind
of Dirac-like Hamiltonians be also consistent with some symmetric interactions with external 
fields
\cite{visinescu04,breev16}.

 \section*{Acknowledgments}

This work was partially supported by Junta de Castilla y Le\'on (BU229P18) and  by Ankara University BAP No. 20L0430005. D. Demir K{\i}z{\i}l{\i}rmak acknowledges Ankara Medipol University.  





\end{document}